\documentclass
  [ sigconf
  , screen
  ]{acmart}

\setcopyright{none}
\usepackage{enumitem}
\usepackage{xspace}
\usepackage{xargs}
\usepackage{subcaption}
\usepackage{hyperref}
\usepackage{pifont}
\usepackage{graphics}
\usepackage{tikz}

\usepackage[plain]{algorithm}
\usetikzlibrary{fit,positioning}
\usepackage{pgfplots}
\pgfplotsset{compat=1.18}
\usepackage{threeparttable}
\usepackage{tcolorbox}
\tcbuselibrary{skins}
\usepackage{listings}
\newcommand{\appname}{\textsc{Emerge}\xspace}
\newcommand{\op}[1]{\mathsf{#1}}
\newcommand{\opbox}[2][]{%
\tikz[baseline=(X.base)]{
  \node[draw, rounded corners, fill=#1!20, inner xsep=4pt, inner ysep=0pt] (X) {\strut #2};
}}
\DeclareRobustCommand{\detLzero}{%
  \tikz[baseline=-0.55ex]\fill (0,0) circle (0.75ex);%
}
\DeclareRobustCommand{\detLone}{%
  \tikz[baseline=-0.55ex]{%
    \draw (0,0) circle (0.75ex);%
    \begin{scope}
      \clip (0,0) circle (0.75ex);
      \fill (-0.75ex,-0.75ex) rectangle (0,0.75ex);
    \end{scope}
  }%
}
\DeclareRobustCommand{\detLtwo}{%
  \tikz[baseline=-0.55ex]\draw (0,0) circle (0.75ex);%
}
\DeclareRobustCommand{\detMiss}{\ding{55}}

\author{Qi Zhan}
\orcid{0000-0002-6800-1857}
\affiliation{%
  \institution{The State Key Laboratory of Blockchain and Data Security, Zhejiang University}
   \city{Hangzhou}
  \country{China}
}
\email{qizhan@zju.edu.cn}

\author{Xing Hu}
\orcid{0000-0003-0093-3292}
\authornote{Corresponding Author}
\affiliation{
  \institution{The State Key Laboratory of Blockchain and Data Security, Zhejiang University}
   \city{Hangzhou}
  \country{China}
  }
\email{xinghu@zju.edu.cn}

\author{Xin Xia}
\orcid{0000-0002-6302-3256}
\affiliation{
  \institution{The State Key Laboratory of Blockchain and Data Security, Zhejiang University}
   \city{Hangzhou}
  \country{China}
}
\email{xin.xia@acm.org}

\author{Shanping Li}
\orcid{0000-0003-2615-9792}
\affiliation{
  \institution{The State Key Laboratory of Blockchain and Data Security, Zhejiang University}
   \city{Hangzhou}
  \country{China}
}
\email{shan@zju.edu.cn}

\title{Verify Implementation Equivalence of Large Models}

\begin{abstract}

  Verifying whether two implementations of the same large model are equivalent across frameworks is difficult in practice. Even when they realize the same
  computation, their graphs may differ substantially in operator decomposition, tensor layout, and the use of fused or opaque kernels, making manual rewrite rules
  hard to build and maintain.
  We present \appname, a framework for checking \emph{Implementation Equivalence} over computation graphs of large-model implementations. Instead of writing rules manually, \appname represents the two implementations in an e-graph, infers candidate relations from execution values, and synthesizes
  rewrite rules on demand when existing rules are insufficient. Each synthesized rule is validated using the strongest applicable method, including SMT-based
  checking for symbolically tractable cases and constraint-aware randomized testing for opaque kernels, and then propagated through e-graph rebuilding to
  establish larger equivalences.
  Our current implementation targets inference computation graphs captured from HuggingFace Transformers and vLLM. Our evaluation shows that \appname establishes equivalence for correct
  implementation pairs at practical cost, while also providing useful by-products for debugging: it detects 10 of 13 known implementation bugs, uncovers 8 previously unknown implementation issues that were later confirmed by developers. In addition, \appname synthesizes high-level
  rules that compare favorably with manually authored ones.
\end{abstract}

\begin{document}
\maketitle

\section{Introduction}

Deep learning deployment today relies on a diverse ecosystem of reference frameworks, compiler stacks, and high-performance serving systems, such as HuggingFace Transformers, XLA, TVM/Ansor, and vLLM~\cite{wolf-etal-2020-transformers,megatroncore,tvm,vllm,Ansor,xla2017}. 
As a result, the same model architecture is often realized by multiple implementations, optimized for serving efficiency or specific hardware targets.
However, these implementations are not guaranteed to be equivalent. Even when derived from the same high-level model specification, they may differ substantially in computation graphs, operator decompositions, tensor layouts, and low-level kernels. We refer to the problem of determining whether such implementations realize the same mathematical function as \emph{Implementation Equivalence}.

When implementation equivalence does not hold, it often leads to \emph{silent bugs}: they do not trigger crashes or exceptions, but instead produce incorrect intermediate tensors or outputs during normal execution~\cite{traincheck}. Such failures have been observed in production ML frameworks, including incorrect loss scaling in Megatron-LM and output mismatches in vLLM caused by implementation bugs in components such as split-QKV handling and RoPE~\cite{megatron_issue673,vllm_issue8017,vllm_issue590,bekman2022bloom}.

Existing research has focused on equivalence checking for deep learning systems, particularly in the setting of \emph{Parallel
Equivalence}~\cite{TTrace,trainverify,wang2025verifydistributeddeeplearning,Aerify}. In this setting, the goal is to verify that
distributed execution strategies, such as tensor sharding or data parallelism, preserve the semantics of a reference implementation.
However, verifying implementation equivalence beyond the parallel setting remains under-explored.
For instance, the HuggingFace Transformers library~\cite{wolf-etal-2020-transformers} provides reference implementations of state-of-the-art models, while vLLM~\cite{vllm} offers highly optimized implementations.
Although these frameworks target the same model architectures, their implementations can differ substantially in computation graphs, tensor layouts, and kernel-level execution.

Recent work has explored equality saturation and e-graphs for proving equivalence in deep learning systems~\cite{Aerify,zulkifli2025verifyingcomputationalgraphsproductiongrade,wang2025verifydistributeddeeplearning}. These approaches
construct a unified representation of computation graphs and iteratively apply human-defined rewrite rules to establish equivalence.
 However, such manually authored rewrite rules are typically tailored to parallel-equivalence transformations and do not transfer easily to cross-framework model implementations.
   Unlike parallel-equivalence settings, where equivalence often follows from a relatively structured set of partitioning and aggregation transformations, cross-framework model implementations introduce two additional challenges:

\textbf{Challenge 1: Divergent Implementation Paths.}
Even when mathematically equivalent, two implementations may use different operator decompositions, execution plan, and tensor layouts~\cite{TASO, mirage}. For example, GPT-2~\cite{gpt2}'s linear projection is realized as $\op{addmm}$ in the Transformers but as $\op{linear}$ in vLLM, with the weight tensor stored in a transposed layout and the surrounding execution plan organized differently. As a result, the corresponding computation graphs can differ substantially in structure, making it impractical to manually encode rewrite rules for every such divergence.

\textbf{Challenge 2: Opaque Semantics of Custom Kernels.}
Custom kernels make equivalence checking difficult because their semantics are often not explicit at the computation-graph level.
To maximize inference-time performance, modern frameworks rely heavily on such operators, often implemented as fused CUDA kernels or hardware-specific primitives~\cite{triton, flashattention,flashattention3}.
These operators are difficult to model precisely with existing verification techniques.
In addition, they can introduce small numerical differences that accumulate along the computation graph~\cite{float}, making direct output-based testing unreliable for establishing model equivalence. Manually writing rules for such operators requires significant domain expertise and does not generalize well across frameworks.

\begin{figure*}[!t]
  \centering
  \includegraphics[width=\linewidth]{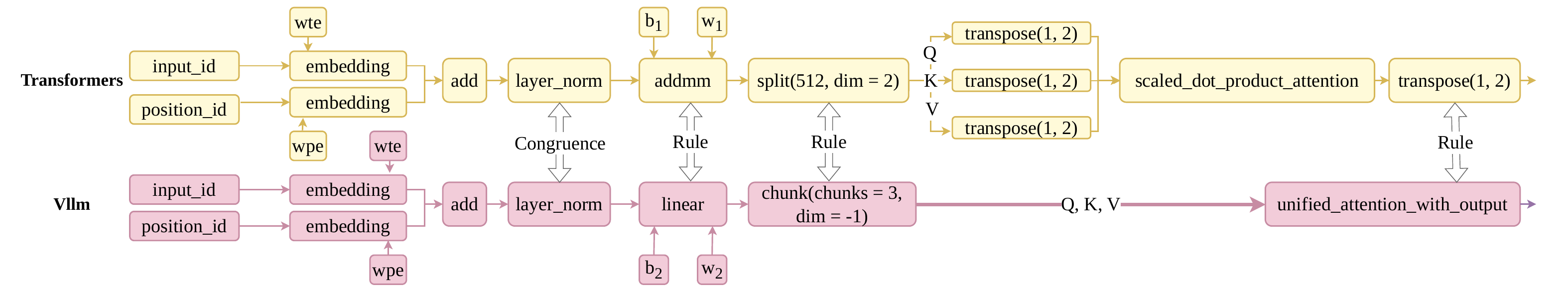}
  \caption{A part of GPT-2 Model used to illustrate equivalence verification between two implementations. Simplified and adjusted for clarity.}
  \Description{A side-by-side fragment of two GPT-2 computation graphs, one from Transformers and one from vLLM, showing matched embedding and normalization prefixes, divergent operator choices such as addmm versus linear, and an attention path connected by congruence and synthesized rules.}
  \label{fig:nutshell}
\end{figure*}

To address these challenges, we propose \appname, a framework for checking Implementation Equivalence over computation graphs of large-model implementations.
The key idea is to verify equivalence without relying on hand-written rewrite rules: instead, \appname discovers the relations needed to connect different implementations directly from their executions and synthesizes rewrite rules on the fly.

  Our approach represents both implementations in a joint e-graph~\cite{egg} and incrementally establishes equivalence between their nodes.
  For each candidate relation, \appname first attempts to establish it using available rules; only when this fails does it synthesize a new rule
  by abstracting over shared dependencies and validate it with SMT solving or constraint-aware randomized testing. Each newly established
  equivalence is then propagated through e-graph rebuilding. This design allows \appname to handle divergent implementation paths,
  framework-specific operator fusions, and opaque kernels within a unified verification framework.

  We implement \appname on computation graphs extracted by TorchDynamo~\cite{torchfx} and evaluate it on model pairs drawn from Transformers and vLLM. Our evaluation shows that \appname is effective along three dimensions. First, it detects real implementation bugs, including 10 of 13 known cases and 8 previously unknown implementation issues, while providing useful fault localization signals. Second, it establishes equivalence for correct model pairs at practical cost. Third, it synthesizes block-level rewrite rules that compare favorably with manually authored ones.

\noindent \textbf{Contributions.} Our main contributions are as follows:
\begin{itemize}[leftmargin=*]
\item \textbf{Problem Formalization.} We formulate \emph{Implementation Equivalence} as a distinct verification problem for multiple implementations of the same model,
  together with a formal framework over joint graphs, e-classes, and validated rewrite rules.
\item \textbf{Dynamic Rule Synthesis.} We propose \appname, which automatically discovers and validates rewrite rules on the fly, without manual rule writing.
\item \textbf{Evaluation.} We evaluate \appname on model pairs drawn from Transformers and vLLM, showing that it detects 10 of 13 known bugs, identifies 8 previously unknown implementation issues, provides useful fault localization, and establishes equivalence for correct implementation pairs at practical cost.
\end{itemize}

The rest of the paper is organized as follows. In \S~\ref{sec:nutshell}, we provide an overview of \appname through an illustrative example. Section~\ref{sec:approach} presents the approach, and Section~\ref{sec:implementation} describes implementation details. We present evaluation results in Section~\ref{sec:evaluation}, discuss threats to validity and limitations in Section~\ref{sec:discussion}, review related work in Section~\ref{sec:related}, and conclude in Section~\ref{sec:conclusion}.

\section{\appname in a Nutshell}\label{sec:nutshell}

In this section, we provide a high-level overview of \appname through an illustrative example, shown in Figure~\ref{fig:nutshell}.
We consider two implementations of the same GPT-2 subcomponent: one from HuggingFace Transformers, the other from vLLM. Although mathematically equivalent, they differ substantially in operator choices, computation graph structure, and tensor layouts. 

\paragraph{Input Relation.}
We begin by establishing equivalences between the leaf nodes (inputs and parameters) of the two computation graphs. Since both implementations take the same token IDs and share embedding weights, most input nodes can be matched directly.
A more interesting case arises for the weight tensors \opbox[yellow]{$w_1$} and \opbox[purple]{$w_2$}. These represent the same logical matrix but are stored in transposed layouts by the two frameworks. By comparing the loaded weight values, \appname automatically infers the relation $w_1 \Longleftrightarrow \op{transpose}(w_2).$

\paragraph{Congruence and Graph Rebuilding.}
After aligning input nodes, \appname propagates these relations to downstream operators through \emph{congruence}: if $a \equiv
  b$, then $f(a) \equiv f(b)$ whenever the same operator $f$ is applied in the same position.
In this example, the two implementations share the same prefix computations, including the embedding, addition, and layer normalization blocks.
Once their inputs are aligned, congruence automatically merges the corresponding  intermediate nodes without requiring any rules.
The \opbox[yellow]{layernorm} and \opbox[purple]{layernorm} nodes in two implementations are merged directly through congruence.
  After each successful merge, \appname rebuilds the e-graph to propagate the new equivalence; the precise procedure is described
  in~\S\ref{sec:node-merge}.

\paragraph{Rule Synthesis and Application.}
  After rebuilding, \appname searches for node pairs that remain in different equivalence classes but have matching concrete values. Such pairs suggest candidate relations not yet captured by the current rule set.
For the pair \opbox[yellow]{addmm} and \opbox[purple]{linear}, \appname synthesizes a candidate rewrite rule by abstracting over the concrete arguments: \[ \op{linear}(a, \op{transpose}(c), b) \Longleftrightarrow \op{addmm}(b, a, c).  \]
The rule is then \emph{validated} via randomized testing (\S\ref{sec:rule-validation}). Once validated, it is added to the rule set and immediately applied to merge the two nodes into the same equivalence class. 
 This merge is then propagated to downstream operators through rebuilding, allowing \appname to lift the newly established equivalence through the rest of the graph~(\S\ref{sec:node-merge}).

Following the same procedure, \appname infers a rule relating \opbox[yellow]{split} and \opbox[purple]{chunk}:
$
\op{split}(a, 512, \text{dim}=2)
\;\Longleftrightarrow\;
\op{chunk}(a, \text{chunks}=3, \text{dim}=-1),
$
which holds only when the shape of $a$ satisfies $[i, 3\times 512]$, where $i$ denotes the number of input tokens.  
This precondition is necessary because the correspondence between $\op{split}$ and $\op{chunk}$ depends on evenly divisible dimensions. 
\appname validates the rule formally via SMT solving under the shape constraints~(\S\ref{sec:rule-validation}).

\paragraph{Opaque Kernels.}
Finally, we consider the attention computation~\cite{attention},
where the two implementations differ most substantially. 
Transformers decomposes attention into separate operators via scaled dot-product attention and output projection, whereas vLLM fuses the same computation into a single \texttt{unified\_attention} kernel.
\appname infers the following equivalence between \\
 \opbox[yellow]{scaled\_dot\_product\_attention} and \opbox[purple]{unified\_attention}:
\[
\begin{aligned}
& \op{unified\_attention}(Q, K, V)
\;\Longleftrightarrow\;&\; \\
& \op{transpose}\left(
\op{scaled\_dot\_product\_attention}\left(
\begin{array}{l}
\op{transpose}(Q),\\
\op{transpose}(K),\\
\op{transpose}(V)
\end{array}
\right)
\right).
\end{aligned}
\]
Here we abbreviate $\op{transpose}(a, 1, 2)$ as $\op{transpose}(a)$ for clarity.
Since this kernel is opaque to symbolic reasoning,  
\appname validates the rule via randomized testing~(\S\ref{sec:rule-validation}).  
The complete rule set and validation verdicts  
are made available for manual inspection.  

At this point, the output nodes belong to the same equivalence class, establishing the implementation equivalence of the two graphs. 
Although this walkthrough covers only a fragment of GPT-2,  
the synthesized rules are immediately reusable across later layers, amortizing the synthesis cost across the entire model.  
  This example illustrates the core intuition behind \appname, but leaves out several important design details, including candidate
  generation, rule synthesis and validation, and e-graph maintenance. We now present the full framework.

  \begin{algorithm}[t]
  \caption{High-level Algorithm of \appname.}
  \label{alg:approach}
\begin{lstlisting}[  
  language=Python,  
  gobble=4,  
  basicstyle=\ttfamily,  
  numbers=none,  
  frame=single,
  keywords={def, for, if, else, while, return, not, in},  
  keywordstyle=\bfseries,  
  commentstyle=\color{gray},  
  escapechar=`,  
  literate=  
    {EMPTY}{$\emptyset$}{1}  
    {UNION}{$\cup$}{1},  
]  
    def compare(P, Q):
      G = initialize_egraph(P, Q) # build joint e-graph
      R = EMPTY  
      for i in range(MAX_ITERATIONS):
        # may augment G with auxiliary nodes
        for (u, v) in find_possible_relations(G): # `\S\ref{sec:relation-inference}`  
          proved = apply_rules(R, u, v) # `\S\ref{sec:rule-apply}`  
          if not proved: # otherwise, synthesize new rule  
            rule = synthesize_rule(G, u, v) # `\S\ref{sec:rule-synthesis}`  
            if rule and validate(rule):  # `\S\ref{sec:rule-validation}`  
              R = R UNION {rule}  
              proved = True  
          if proved:  
            G.merge(u, v)  
            G.rebuild()  # `\S\ref{sec:node-merge}`  
        if outputs_equivalent(G):
          return EQUIVALENT
      return FAILED, localize(G) # `\S\ref{sec:localization}`
\end{lstlisting}  
\end{algorithm}

\section{Approach}\label{sec:approach}

\appname verifies the functional equivalence of two computation graphs through an iterative procedure on a joint e-graph. The overall workflow is illustrated in Algorithm~\ref{alg:approach}. 
Given two computation graphs, \appname repeatedly proposes candidate relations between nodes (\S\ref{sec:relation-inference}) and attempts to establish them using existing rules (\S\ref{sec:rule-apply}). When no existing rule applies, it synthesizes a candidate rule from the current e-graph state (\S\ref{sec:rule-synthesis}) and validates the rule (\S\ref{sec:rule-validation}) before merging the corresponding nodes. After each successful merge, the e-graph is rebuilt to maintain congruence and refresh attached execution data (\S\ref{sec:node-merge}).
At the end of each iteration, \appname checks whether the output nodes $o_1$ and $o_2$ belong to the same equivalence class. If so, the two graphs are declared \textit{equivalent}. Otherwise, if a predefined iteration limit is reached first, the verification attempt fails and \appname invokes failure localization to identify the first pair of mismatched nodes that best explains the remaining divergence.  

\subsection{Formalization}
We begin by formalizing the key components of our approach. 

\emph{Computation Graph.}
A computation graph is represented as a directed acyclic graph (DAG) $G = (V, E)$, where each node $v \in V$ represents either an operation (e.g., addition, multiplication, convolution) or an input tensor and each edge $(u, v) \in E$ indicates that the output of node $u$ is an input to node $v$.
Given two graphs $G_1 = (V_1, E_1)$ and $G_2 = (V_2, E_2)$, we construct a joint graph $G = (V, E)$ where $V = V_1 \cup V_2$ and $E = E_1 \cup E_2$. 

  \emph{Equivalence Relation and E-Classes.}
  Implementation Equivalence is a graph-level semantic notion: it asks whether two implementations realize the same function. During verification, however, \appname
  maintains a node-level equivalence relation $\equiv$ over the nodes of the joint graph $G$. This relation induces a partition of $V$ into equivalence classes, or
  \emph{e-classes}; for a node $n \in V$, we write $\overline{n}$ for the e-class containing $n$. Intuitively, nodes in the same e-class are treated as equivalent by
  the e-graph. Each e-class is associated with concrete execution data obtained from dynamic execution; we denote this attachment by a partial function $f :
  \overline{V} \to D$, where $D$ is the domain of runtime tensor values.
  Throughout the paper, we use plain symbols such as $n,u,v$ for nodes and overlined symbols such as $\overline{n}, \overline{u}, \overline{v}$ for their corresponding e-classes.


\emph{Rewrite Rules.}
A rewrite rule is an equivalence of the form
$
l \;\Longleftrightarrow\; r,
$
where $l$ and $r$ are operator terms built from the same operator vocabulary as the computation graph. An operator term is either a
variable, a literal constant or an operator applied to a finite number of sub-terms. We write $\mathrm{FV}(t)$ for the free variables of a term $t$.
A substitution $\sigma$ maps the free variables of a rule to e-classes.
Intuitively, a rewrite rule describes two parameterized computation subgraphs that produce the same result under all consistent
substitutions of their free variables. In our system, rewrite rules are treated symmetrically, so either side may be matched and used to establish equivalence.

\emph{Maintained Structures.}
During verification, \appname maintains the joint graph $G$, the current equivalence relation $\equiv$, the attached execution data $f$, and the set $\mathcal{R}$ of
validated rewrite rules. The rule set $\mathcal{R}$ may be initialized either as empty or with a seed set of manually authored rules. These components are updated iteratively as new relations are inferred, rules are synthesized and validated, and e-classes are merged and
rebuilt. The procedure succeeds when the designated output nodes $o_1 \in V_1$ and $o_2 \in V_2$ are placed in the same e-class.\footnote{Without loss of generality,
we assume single output nodes; in practice, both graphs may have multiple outputs, and we require pairwise equivalence.}



The following sections use the running example in Figure~\ref{fig:rulelearn}. In the figure, dashed boxes denote e-classes and solid boxes denote nodes. One subgraph
computes a linear projection through the decomposed pattern $\op{add}(\op{mm}(i_1, w_1), b_1)$, while the other uses the fused operator $\op{linear}(i_2, w_2, b_2)$.
Although the two forms differ in operator structure and weight layout, they correspond to the same logical computation. We use this example to show how \appname
aligns inputs, synthesizes rewrite rules, and eventually merges the two outputs into the same e-class.

\subsection{Relation Inference}
\label{sec:relation-inference}

\begin{figure}[!t]
  \centering
  \includegraphics[width=\linewidth]{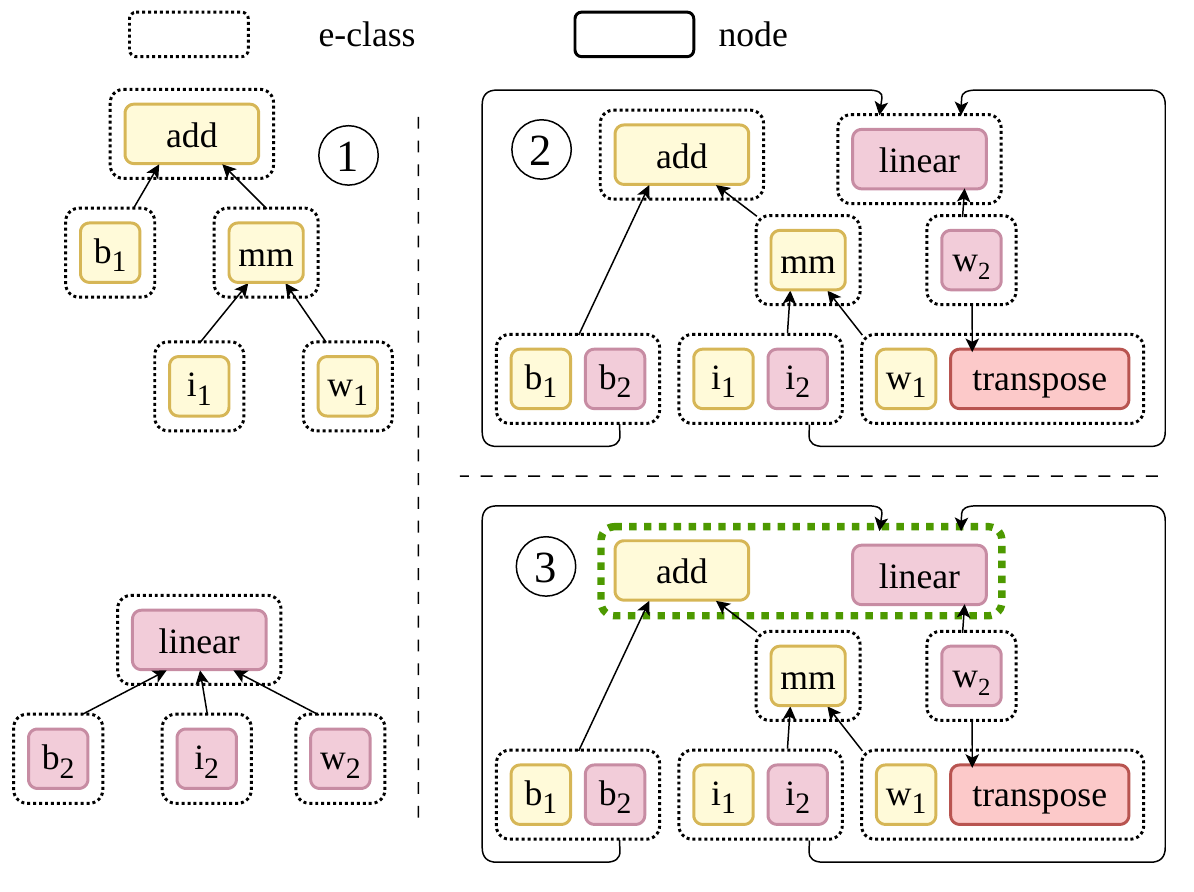}
  \caption{Rule synthesis from execution traces. 
  \ding{172} Initial relation \ding{173} Relation inferred from input values \ding{174} Relation inferred from rule synthesis. }
  \Description{An e-graph rule-synthesis example shown in three stages: an initial relation between add and linear subgraphs, a new relation inferred from matching input values after aligning transformed weights, and a final relation justified by a synthesized rewrite rule that merges the outputs.}
  \label{fig:rulelearn}
\end{figure}

\appname infers candidate relations by comparing the concrete tensor values attached to e-classes.
For two e-classes $\overline{u}$ and $\overline{v}$, if $f(\overline{u})$ and $f(\overline{v})$ match up to a configurable numerical tolerance $\epsilon$, then
$(\overline{u}, \overline{v})$ is treated as a candidate relation. This direct comparison is effective in many cases, but it is not sufficient on its own
because semantically equivalent tensors may differ in layout, partitioning, or structural organization; for example, one may be the transpose of the other or a
concatenation of its components.
In our implementation, value matching uses an absolute/relative tolerance of $10^{-2}$ by default.
As illustrated in Figure~\ref{fig:rulelearn}, the weight matrices \opbox[yellow]{$w_1$} and \opbox[purple]{$w_2$} are stored in different layouts.
We therefore need to introduce an auxiliary node to relate them.

To handle such cases, \appname uses a small \emph{transformation grammar} to describe value-level operations that relate semantically equivalent tensors with different layouts or structures. The search space is given by the following context-free grammar:
$$
\begin{array}{rcl}
E &\to \op{transpose}(E, \text{axis}_1, \text{axis}_2)  \mid \op{concat}(E, \ldots, E, \text{axis}) \\ &\mid \op{reshape}(E, \text{shapes}) \mid \op{split}(E, \text{size}, \text{axis}) \\
\end{array}
$$
  This grammar captures common layout and structural transformations in deep learning frameworks, including reordering, reshaping, and partitioning. \appname then
  performs bottom-up enumerative synthesis over this grammar to search for an expression that transforms one tensor value into the other. When such an expression
  is found, the corresponding auxiliary nodes are added to the joint graph. Moreover, these transformations admit natural inverses under compatible parameters
  (e.g., $\op{concat}$ and $\op{split}$ along the same axis), so discovering a transformation from $a$ to $b$ also yields the reverse relation from $b$ to $a$,
  and it suffices to search in one direction.

  Once such a transformation is found, \appname adds the corresponding auxiliary nodes to the joint graph and uses them to relate the original nodes. If both
  original nodes are leaves, such as model inputs or parameters, the transformed node can be merged directly with its counterpart to establish an initial
  equivalence. In Figure~\ref{fig:rulelearn}, for example, \appname introduces the auxiliary node \opbox[red]{transpose} on top of \opbox[purple]{$w_2$} and
  merges it with \opbox[yellow]{$w_1$} in~\ding{173}. If the matched nodes are non-leaf nodes, however, the added auxiliary structure only exposes a candidate
  equivalence. \appname must still justify it through rule application~(\S\ref{sec:rule-apply}) or, when no existing rule applies, through rule
  synthesis~(\S\ref{sec:rule-synthesis}).

\subsection{Rule Synthesis}
\label{sec:rule-synthesis}

When \appname identifies a candidate relation between e-classes $\overline{u}$ and $\overline{v}$, it first attempts to establish it using existing rules in $\mathcal{R}$.
If no existing rule applies, \appname then synthesizes a rule using representative nodes $u \in \overline{u}$ and $v \in \overline{v}$.
Concretely, rule synthesis seeks an equivalence $l \Longleftrightarrow r$ such that $l$ matches $u$ and $r$ matches $v$ in the current e-graph. However, not every pair of target e-classes admits such a rule. We therefore first determine whether the two sides can be abstracted over a common frontier.

To this end, we introduce the notion of an \emph{abstraction frontier} of an equivalence class. Intuitively, a frontier is a set of e-classes whose values jointly determine the value of the target e-class. Coarser frontiers stop earlier and yield more general abstractions, while finer frontiers expand deeper and preserve more internal structure.
Formally, for an equivalence class $\overline{n}$, we write
$\mathcal{C}(\overline{n}) \subseteq \mathcal{P}(\overline{V})$ for the set of candidate abstraction frontiers of $\overline{n}$.
For a representative node $n \in \overline{n}$, let $\mathcal{C}(n)$ denote the frontier family obtained by expanding through $n$.
Then $\mathcal{C}(n)$ and $\mathcal{C}(\overline{n})$ can be computed recursively as
\begin{align}  
\mathcal{C}(n) &=  
\begin{cases}  
\bigl\{\{\overline{n}\}\bigr\},  
  & n \text{ is a leaf}, \\[6pt]  
\bigl\{\{\overline{n}\}\bigr\}  
  \cup  
  \bigl\{\,  
    {\textstyle\bigcup_{i=1}^{k}} S_i  
  \;\bigm|\;  
    \vec{S} \in  
    {\textstyle\prod_{i=1}^{k}}  
      \mathcal{C}(\overline{l_i})  
  \,\bigr\},  
  & \text{otherwise},  
\end{cases}  
\label{eq:frontier-node}  
\\[4pt]  
\mathcal{C}(\overline{n}) &=  
  \textstyle\bigcup_{m \in \overline{n}} \mathcal{C}(m),  
\label{eq:frontier-class}  
\end{align}  
where $\overline{l_1},\ldots,\overline{l_k}$ are the children of~$n$  
and $\vec{S} = (S_1,\ldots,S_k)$.  
Intuitively, frontier enumeration may either stop at the current e-class or continue expanding through its children; taking the union over all representatives
  allows enumeration to proceed through any equivalent form already present in the e-graph.
    For example, in Figure~\ref{fig:rulelearn},
  $\{\overline{\op{add}}\} \in \mathcal{C}(\overline{\op{add}})$,
  $\{\overline{\op{mm}}, \overline{b_1}\} \in \mathcal{C}(\overline{\op{add}})$,
  and
  $\{\overline{b_1}, \overline{w_1}, \overline{i_1}\} \in \mathcal{C}(\overline{\op{add}})$.

Two equivalence classes $\overline{u}$ and $\overline{v}$ can admit a rewrite rule only if they share a common frontier, i.e.,
$\mathcal{C}(\overline{u}) \cap \mathcal{C}(\overline{v}) \neq \emptyset$.
If their frontier families are disjoint, then no meaningful symbolic relation can be synthesized between them. In the running example, after input alignment we have $\overline{b_1}=\overline{b_2}$ and $\overline{i_1}=\overline{i_2}$, so the \opbox[yellow]{add} and \opbox[purple]{linear} nodes share the common frontier $\{\overline{b_1}, \overline{i_1}, \overline{w_2}\}$.
By contrast, the \opbox[yellow]{mm} node and the \opbox[purple]{linear} node do not share any common frontier, and hence no rule is synthesized between them.

A key design goal is not merely to synthesize a valid rule, but to synthesize the \emph{most general} one.
When multiple common frontiers are available, they induce
symbolic patterns at different abstraction levels. If frontier enumeration stops too early, the resulting rule preserves unnecessary internal structure and becomes
overly specific.
For example, a naive synthesis procedure might stop at the e-class containing $\op{layernorm}(a)$ and produce the overly specific rule
$\op{linear}(\op{layernorm}(a), \op{transpose}(c), b) \Leftrightarrow \op{addmm}(b, \op{layernorm}(a), c)$,
which unnecessarily constrains the first argument to be a $\op{layernorm}$ output.
By continuing the abstraction one step further and treating its input as a symbolic variable, \appname instead synthesizes the more general rule
$\op{linear}(a, \op{transpose}(c), b) \Leftrightarrow \op{addmm}(b, a, c)$.

Equipped with the above definitions, we now present the algorithm for rule synthesis, shown in Algorithm~\ref{alg:rulesynthesis}. For each target e-class, the
algorithm enumerates a sequence of frontier-pattern pairs. Each pair $(S, p)$ consists of a frontier $S$ and a symbolic pattern $p$ obtained by abstracting the target
over $S$.
Here $\Pi$ denotes Cartesian-product enumeration.
The enumeration proceeds in a top-down manner, from the coarsest abstraction to progressively finer ones. Starting from a target e-class $\overline{n}$, the
algorithm first yields the frontier $\{\overline{n}\}$ (line~9), treating the entire e-class as a single symbolic atom. It then recursively expands
representative nodes toward their inputs (lines~12--26), refining the current frontier and constructing correspondingly more specific symbolic patterns. In this way, frontier-pattern pairs are generated in decreasing order of generality.
Given the two enumerated sets $P_u$ and $P_v$, the procedure searches over $\Pi(P_u, P_v)$ for a pair $(S_u, p_u) \in P_u$ and $(S_v, p_v) \in P_v$ such that
$S_u = S_v$ (lines~4--7). Because frontier enumeration emits frontier-pattern pairs from coarser to finer frontiers, the first such match corresponds to the coarsest shared abstraction frontier between $\overline{u}$ and $\overline{v}$, and therefore yields the most general candidate rule under the current e-graph:
$p_u \;\Longleftrightarrow\; p_v.$
If no such pair exists, the algorithm concludes that no rule can be synthesized for $\overline{u}$ and $\overline{v}$ under the current graph state.

In the running example, the extracted symbolic pattern summarizes the decomposed subgraph $\op{add}(\op{mm}(i_1, w_1), b_1)$ as
the compound term $\op{addmm}(b, a, c)$. In the end, \appname synthesizes the equivalence
$\op{linear}(a, \op{transpose}(c), b) \Longleftrightarrow \op{addmm}(b, a, c)$,
thereby merging the \opbox[yellow]{add} root and the \opbox[purple]{linear} root in \ding{174} of Figure~\ref{fig:rulelearn}.

    \begin{algorithm}[!t]
  \begin{lstlisting}[language=Python, gobble=2, numbers=left,
    keywords={def, for, if, in, return, continue, yield}, frame=single, numbersep=-8pt, basicstyle=\ttfamily, escapechar=`]
      def synthesize_rule(egraph, `$\overline{u}$`, `$\overline{v}$`):
        `$P_u$` = enumerate_frontier_patterns(egraph, `$\overline{u}$`)
        `$P_v$` = enumerate_frontier_patterns(egraph, `$\overline{v}$`)
        for ((`$S_u$`, `$p_u$`), (`$S_v$`, `$p_v$`)) in `$\Pi$`(`$P_u$`, `$P_v$`):
          if `$S_u$` == `$S_v$`: # depend on the same frontier                    
              return Rule(`$p_u$`, `$p_v$`)

      def enumerate_frontier_patterns(egraph, `$\overline{n}$`):
        yield (`$\{\overline{n}\}$`, Atom(`$\overline{n}$`))
        # Expand through operators in the same e-class
        for n in `$\overline{n}$`:
            if n.op == "input":
                continue
            child_patterns = [
                enumerate_frontier_patterns(egraph, arg)
                for arg in n.args
            ]
            # Cartesian product over children
            for combo in `$\Pi$`(*child_patterns):
                `$S$` = set()
                args = []
                for (`$S_i$`, `$p_i$`) in combo:
                    `$S$` = `$S$` `$\cup$` `$S_i$`
                    args.append(`$p_i$`)
                `$p$` = Node(n.op, args)
                yield (`$S$`, `$p$`)

  \end{lstlisting}
    \caption{Rule synthesis by shared frontier enumeration.}
    \label{alg:rulesynthesis}
  \end{algorithm}

\subsection{Rule Validation} \label{sec:rule-validation}

Synthesized rules are not admitted into $\mathcal{R}$ immediately. Before a rule can participate in equality saturation, \appname validates it using the
strongest applicable method. The validation pipeline serves two purposes: to reject ill-formed or semantically unsound rules, and to attach an evidence level to
each accepted rule.
\textbf{Pre-checking and Rule Classification.} Validation begins with a lightweight structural pre-check. In this phase, \appname rejects malformed or
unsupported rules, such as rules with inconsistent operator attributes. Only rules
that pass this pre-check proceed to semantic validation.

The remaining rules are then classified according to their semantic structure in order to select the strongest applicable validation strategy. At a high level,
we distinguish three classes:
(i) \textit{Scalar-Logic}, containing only arithmetic, boolean, or bitwise operators;
(ii) \textit{Tensor-Rearrangement}, consisting of structural tensor transformations such as \texttt{view}, \texttt{transpose}, \texttt{reshape}, and
\texttt{split}; and
(iii) \textit{Opaque-Heavy}, containing fused kernels, black-box operators, or numerically complex tensor operators such as \texttt{embedding},
\texttt{softmax}, and \texttt{linear}.

\textbf{Symbolic Verification via SMT.} For rules whose semantics can be modeled precisely, \appname invokes an SMT solver to prove equivalence under the
inferred constraints.
\begin{itemize}[leftmargin=*]
  \item \textbf{Scalar Arithmetic.} Expressions are lowered to SMT real and boolean terms. The solver checks the negation of the candidate equivalence; if the
negation is unsatisfiable, the rule is proved correct and marked \emph{formally verified}.
  \item \textbf{Tensor Rearrangement.} For structural tensor rules, \appname models each tensor as a symbolic array whose elements are distinct symbolic
variables. Operators such as \texttt{transpose}, \texttt{reshape}, and \texttt{split}/\texttt{concat} are lowered to index transformations. Under satisfiable
shape constraints, the solver proves element-wise equality between the two sides.
\end{itemize}

\textbf{Constraint-Aware Randomized Testing.} For rules involving fused kernels, numerically complex operators, or operators outside the supported SMT,
\appname falls back to empirical validation. To make randomized testing meaningful, the validator first solves the rule's preconditions and then samples inputs
that satisfy them.
\begin{itemize}[leftmargin=*]
  \item \textbf{Shape-Aware Generation.} \appname solves systems of shape constraints, such as \texttt{shape\_dim\_eq} and \texttt{numel\_eq}, to generate valid
tensor shapes satisfying operator requirements such as matrix-multiplication alignment, reshape consistency, or divisibility for chunking and splitting.
  \item \textbf{Domain-Aware Sampling.} For operators with restricted input domains, such as \texttt{embedding}, the validator infers the semantic role of each
tensor (e.g., index tensor versus weight tensor) and samples values from the corresponding admissible domain. This avoids invalid executions such as out-of-
bounds indexing and improves the effectiveness of randomized testing.
\end{itemize}

In addition to shape and domain constraints, the validator also checks operator-specific side conditions, such as index bounds for \texttt{embedding} or
divisibility requirements for \texttt{chunk}. Rules whose inferred preconditions are unsatisfiable are rejected directly, rather than being vacuously accepted.

Finally, \appname assigns each accepted rule a validation level. Rules proved by SMT are labeled \emph{formally verified}, while rules that pass constraint-
aware randomized testing are labeled \emph{empirically validated}. Only rules with one of these accepted validation levels are inserted into $\mathcal{R}$ and
used during subsequent equality saturation.

 \newcommand{\VL}{https://github.com/vllm-project/vllm}
  \newcommand{\HF}{https://github.com/huggingface/transformers}
  \newif\ifnewlink
  \newlinktrue
  \newcommand{\newhref}[2]{\ifnewlink\href{#1}{#2}\else #2\fi}

 \begin{table*}[!ht]
    \begin{threeparttable}
    \caption{Known bug cases used in our evaluation and detection results compared with TTrace.}
    \label{tab:bugs-known}
    \begin{tabular}{ccclll}
    \toprule
    ID & \appname & TTrace & Description & Issue & Fix \\
    \midrule
    1  & \detLone & \detMiss & InternLM2: wrong \texttt{split\_qkv} execution semantics in tensor parallel & \href{\VL/issues/8017}
  {8017} & \href{\VL/pull/8055}{pull/8055} \\
    2  & \detLtwo & \detMiss & OLMo: wrong shard/full weight-loading semantics in tensor parallel & \href{\VL/issues/3775}{3775} &
  \href{\VL/pull/3869}{pull/3869} \\
    3  & \detLone & \detLone & Gemma: MLP GELU approximation mismatch (\texttt{tanh} vs exact) & \href{\VL/issues/3411}{3411} &
  \href{\VL/pull/3653}{pull/3653} \\
    4  & \detLone & \detLone & GPT-J: incorrect RoPE rotation scheme & \href{\VL/issues/590}{590} & \href{\VL/pull/941}{pull/941} \\
    5  & \detLtwo & \detLone & GLM: rotary / attention path mismatch & \href{\VL/issues/16904}{16904} & \href{\VL/pull/16912}
  {pull/16912} \\
    6  & \detLzero & \detLtwo & Command-R: wrong RoPE base parameter priority & \href{\VL/issues/3892}{3892} & \href{\VL/pull/3919}
  {pull/3919} \\
    7  & \detLzero & \detLone & Gemma2: context/window boundary exposed as rotary/cache inconsistency & \href{\VL/issues/6220}{6220} &
  \href{\VL/pull/10584}{pull/10584} \\
    8  & \detLzero & \detLtwo & Qwen2-VL: weights mapper / parameter-path mismatch & \href{\VL/issues/18976}{18976} & \href{\VL/
  pull/19054}{pull/19054} \\
    9  & \detLzero & \detMiss & Phi: declared max position exceeds materialized rotary cache size & \href{\VL/issues/9502}{9502} &
  \href{\VL/pull/9503}{pull/9503} \\
    10 & \detLone & \detLone & Llama4-Vision: wrong vision rotary cache representation / dtype path & \href{\VL/issues/25888}{25888} &
  \href{\VL/pull/25889}{pull/25889} \\
    11 & \detMiss & \detMiss & Mistral long prompt / sliding-window behavior; semantics are hidden & \href{\VL/issues/2059}{2059} &
  \href{\VL/pull/2088}{pull/2088} \\
    12 & \detMiss & \detMiss & Phi-3.5: LongRoPE branch / KV-cache semantics hidden by data-dependent branching & \href{\VL/
  issues/27414}{27414} & \href{\VL/pull/27431}{pull/27431} \\
    13 & \detMiss & \detMiss & GGUF: TP embedding/materialization bug not isolated under current method & \href{\VL/issues/7880}{7880}
  & \href{\VL/pull/7954}{pull/7954} \\
    \bottomrule
    \end{tabular}
    \begin{tablenotes}[flushleft]
    \footnotesize
    \item[] \textbf{Legend (Detected):} \detLzero~exact bug root; \detLone~same functional block; \detLtwo~farther from the root;
  \detMiss~missed.
    \end{tablenotes}
    \end{threeparttable}
  \end{table*}

\subsection{Rule Application}  
\label{sec:rule-apply}  
  
A validated rewrite rule is applied only after relation inference has identified a candidate pair of nodes. Rule application in \appname is therefore candidate-driven: \appname first proposes a candidate relation and then checks whether an existing rule can justify it. Unlike conventional term rewriting, matching in \appname is \emph{e-class aware}: because an e-class may contain multiple nodes, a rule need not match a single fixed node form. Instead, it may match through any representative already contained in the e-classes.
  
Formally, given a rule $l \Longleftrightarrow r$, \appname searches for a substitution
$\sigma : \mathrm{FV}(l)\cup \mathrm{FV}(r) \to \overline{V}$
from rule variables to e-classes. Literal constants and operator parameters are matched syntactically as part of the term structure, rather than through $\sigma$. We say that the rule matches a pair of e-classes $(\overline{u}, \overline{v})$ if there exist representatives $u \in
\overline{u}$ and $v \in \overline{v}$ such that the left-hand side $l$ matches $u$ under $\sigma$ and the right-hand side $r$ matches $v$ under the same $
\sigma$. Since rules are treated symmetrically in our system, we also accept matches in the reverse direction, i.e., matching $r$ at $u$ and $l$ at $v$.
  
Each matched pair $(u,v)$ provides a concrete witness  
that the rule relates the two e-classes under the  
current bindings.  
\appname merges the corresponding e-classes and invokes  
the rebuilding procedure~(\S\ref{sec:node-merge})  
to restore congruence.  
If a rule carries additional preconditions  
(e.g., shape divisibility constraints inferred  
during synthesis), the match is accepted only when  
these constraints are satisfied by the current  
substitution.

\subsection{E-Class Merge and Graph Rebuild}
  \label{sec:node-merge}

  When two e-classes $\overline{n_1}$ and $\overline{n_2}$ are merged, the new equivalence may induce further merges among their parent nodes through congruence.
  To restore this invariant, \appname performs a worklist-driven rebuilding phase until a fixpoint is reached.

  \begin{enumerate}[leftmargin=*]
    \item \textbf{Canonicalize Parent References.}
    For every parent node $p$ that uses $\overline{n_1}$ or $\overline{n_2}$ as an argument, \appname replaces that argument with the canonical representative of
  the newly merged e-class.

    \item \textbf{Detect New Congruences.}
    After canonicalization, \appname checks whether any parent nodes have become structurally congruent, i.e., they have the same operator and their corresponding
  arguments now refer to the same e-classes. If so, they are merged and added to the worklist for further rebuilding.

    \item \textbf{Refresh Execution Data.}
    The merged e-class retains one attached value as its representative concrete value. Then, for each parent node whose input e-classes have changed, \appname
  re-executes the corresponding operator on the updated representative inputs to obtain a refreshed output value.
  \end{enumerate}

  \subsection{Fault Localization}
    \label{sec:localization}

  When verification terminates without merging the output e-classes, \appname performs failure localization to identify the earliest substantive divergence
  between the two implementations. Rather than reporting only the final unmatched outputs, \appname traces the mismatch backward through the e-graph and returns a
  pair of frontier nodes that best explains why equivalence could not be established.

  The localization procedure is heuristic-guided. Starting from the outputs, it traverses backward along data dependencies while abstracting away nodes treated as
  semantically transparent by the heuristic. It then continues through the primary data-carrying inputs of each operator until it reaches the first frontier whose
  e-class has no cross-graph counterpart. Among all such frontiers, \appname reports the earliest unexplained mismatch.

\section{Implementation}\label{sec:implementation}

\appname is implemented on top of TorchDynamo~\cite{torchfx}, following prior work~\cite{wang2025verifydistributeddeeplearning}, to capture computation graphs from production code (e.g., vLLM and HuggingFace) without manual instrumentation.
To compactly represent and reason about many equivalent expressions, we use an e-graph with Union-Find~\cite{unionfind}, implemented via \texttt{egg}~\cite{egg}. 
Rule validation combines SMT-based checking with Z3~\cite{z3} and
constraint-aware randomized testing, as described in \S\ref{sec:rule-validation}. This section focuses on two practical aspects of the implementation:
specialized input matching and efficiency optimizations.

\subsection{Practical Input Matching}
\label{subsec:input-equiv}

Although input relations can in principle be inferred by the general relation inference procedure in \S\ref{sec:relation-inference}, using that machinery directly on all input pairs is unnecessarily expensive in practice. 
We therefore implement a specialized candidate-generation pipeline for input matching. Given
input sets $I_1$ and $I_2$ from graphs $G_1$ and $G_2$, the goal is to construct a small candidate set $E_{\text{input}} \subseteq I_1 \times I_2$ that is
highly likely to contain the true correspondences.

Purely syntactic alignment by name or argument position is brittle in real systems due to inconsistent naming, layout differences (e.g., channels-first vs. channels-last), and input fusion (e.g., concatenated Q/K/V weights). We therefore treat input matching as candidate generation over concrete values: lightweight heuristics (e.g., shape-compatible weight matching and embedding-table identity) propose a small set of high-probability pairs.
These heuristics are used only to narrow the search space; final acceptance still comes from the core relation inference procedure.

This matching layer is used only for candidate generation. It does not alter the soundness of the verification procedure: every proposed input relation must
still be accepted by the core equivalence engine through value-based relation inference. In this way, the input-matching heuristics improve robustness to inconsistent naming and fused representations while preserving the verification guarantees of the main algorithm.

\subsection{Efficiency Optimizations}\label{subsec:efficiency-opt}
We implement several optimizations to improve efficiency:

\textbf{Dependency-Based Blocking.}  
Two e-classes can be equivalent only if they depend on the same logical inputs. We therefore compute a \emph{dependency signature} for each e-class (the set of reachable input leaf e-classes) and compare candidate relations only when they have identical signatures. This blocks many impossible comparisons early and reduces the search space in practice.

  \textbf{Efficient Frontier Enumeration.}
  Rule synthesis is driven by shared-frontier discovery, so we optimize frontier enumeration in two ways. First, for a candidate relation $(\overline{u},
  \overline{v})$, we precompute the e-classes reachable from both endpoints and restrict enumeration to this intersection, pruning branches that cannot yield a
  shared frontier. Second, Cartesian products in frontier enumeration (Algorithm~\ref{alg:rulesynthesis}, line~20) are generated lazily with Python generators and
  traversed in diagonal order, so that coarser, more general candidates are explored earlier. Together, these optimizations avoid materializing exponentially many
  combinations and allow synthesis to terminate as soon as the first shared frontier is found.

\section{Evaluation}\label{sec:evaluation}

We evaluate \appname on the following research questions (RQs):

\begin{itemize}[leftmargin=*]  
  \item \textbf{RQ1 (Bug Detection):} Can \appname effectively detect and localize real-world implementation bugs?
  \item \textbf{RQ2 (Equivalence \& Scalability):} Can \appname establish equivalence for correct implementation pairs at practical cost?
  \item \textbf{RQ3 (Rule Characterization):}  What kinds of rules does \appname synthesize, and how do they compare with manually authored rules?
\end{itemize}  

\subsection{Experimental Setup}

  \textbf{Baselines.}
  A direct end-to-end baseline is not available for our full setting: existing techniques mainly target tensor-parallel transformations within a shared framework
  and operator semantics, whereas our setting involves cross-framework implementations. We therefore use
  task-specific comparisons: TTrace~\cite{TTrace} serves as the baseline for bug detection in RQ1, and Entangle~\cite{wang2025verifydistributeddeeplearning}
  serves as the baseline for rule characterization in RQ3.

\noindent\textbf{Environment.}
All experiments are conducted on our local server with an Intel Xeon Platinum 8358P CPU, 2\,TB of memory, and one NVIDIA A800 80GB PCIe GPU, running Ubuntu 20.04.6 LTS.
All experiments use a maximum of 2 main verification iterations.

\subsection{RQ1: Bug Detection}\label{subsec:rq1}  

We answer RQ1 through two studies: reproducing known implementation bugs and discovering previously unknown bugs in current releases.
Besides whether \appname detects non-equivalence, we also evaluate localization quality at three levels: \detLzero, where the reported node is the root-cause operator; \detLone, where it falls in the same functional block; and \detLtwo, where it lies further downstream. Here, a functional block denotes the smallest model component implementing a coherent sub-computation, such as rotary embedding, attention, or MLP, as identified from the captured computation graph and the corresponding implementation code.

\subsubsection{Reproducing Known Bugs}

We collected 13 confirmed, independently fixed bugs from the  
issue trackers of vLLM~\cite{vllm} and HuggingFace Transformers~\cite{wolf-etal-2020-transformers}.
For each bug, we reproduce the faulty behavior and  
construct a model pair that exercises the affected code path.  
We then run \appname and record (i)~whether it reports non-equivalence, and (ii)~the first reported mismatch returned by the localization procedure.
For baseline comparison, we run TTrace on the same model pairs under the same configurations and record whether it reports a mismatch and, when available, its localization result.
We report these results in Table~\ref{tab:bugs-known}.

  \begin{table}[!ht]
  \begin{threeparttable}
  \caption{Unknown issues identified by \appname.}
  \label{tab:bugs-new}
  \setlength{\tabcolsep}{3pt}
  \begin{tabular}{ccp{0.54\columnwidth}ll}
  \toprule
  ID & Our & Description & Issue & Fix \\
  \midrule
  14 & \detLzero & GPT-2: \texttt{scale\_attn\_weights} ignored in SDPA path & \newhref{\HF/issues/44380}{44380} & \newhref{\HF/pull/44397}{pull/44397} \\
  15 & \detLzero & GPT-2 family: \texttt{scale\_attn\_weights} mismatch in vLLM & \newhref{\VL/issues/35402}{35402} & \newhref{\VL/pull/35436}{pull/35436} \\
  16 & \detLzero & GPT-BigCode family: \texttt{scale\_attn\_weights} mismatch in vLLM & \newhref{\VL/issues/36565}{36565} & \newhref{\VL/pull/36637}{pull/36637} \\
  17 & \detLtwo & Qwen1: long-context \texttt{logn\_attn} missing in vLLM & \newhref{\VL/issues/36880}{36880} & \newhref{\VL/pull/37089}{pull/37089} \\
  18 & \detLzero & PhiMoE: \texttt{router\_jitter\_noise} hardcoded & \newhref{\VL/issues/35494}{35494} & \newhref{\VL/pull/35499}{pull/35499} \\
  19 & \detLone & OLMoE: \texttt{clip\_qkv} ignored & \newhref{\VL/issues/35513}{35513} & \newhref{\VL/pull/35523}{pull/35523} \\
  20 & \detLtwo & JAIS: learned positional embeddings still trigger ALiBi in vLLM attention & \newhref{\VL/issues/37400}{37400} & \newhref{\VL/pull/37820}{pull/37820} \\
  21 & \detLzero & Phi: \texttt{qk\_layernorm} missing in vLLM & \newhref{\VL/issues/37852}{37852} & \newhref{\VL/pull/37870}{pull/37870} \\
  \bottomrule
  \end{tabular}
  \end{threeparttable}
  \end{table}

\appname detects \textbf{10} of the 13 known bugs. Among the detected cases, 6 are localized exactly (\detLzero), 3 are localized to the same functional block
(\detLone), and 1 is localized further downstream (\detLtwo). For example, Bug~\#4 (GPT-J incorrect RoPE) is localized directly to the
\texttt{rotary\_embedding} operator, while Bug~\#3 (Gemma GELU mismatch) is pinpointed at the \texttt{gelu} node. In these cases, \appname reduces debugging
effort from inspecting the entire model execution to inspecting a single operator and its immediate context.
  Three bugs are missed. Two require data-dependent branches that are not exercised by our current inputs, and one depends on weight-loading semantics outside the
current graph-extraction boundary.

Compared with TTrace, \appname achieves both higher detection coverage and more precise localization on the known-bug benchmark. As shown in Table~\ref{tab:bugs-known}, \appname detects 10 of 13 bugs, whereas TTrace detects 7. Moreover, \appname localizes 6 bugs exactly (\detLzero), while TTrace does
not achieve exact localization on any case.
TTrace localizes bugs by comparing execution traces against a trusted reference under carefully designed numerical tolerances, whereas \appname incrementally establishes local equivalences, merges the corresponding e-classes, and propagates these merges through rebuilding. As a result, already explained differences are removed from subsequent analysis, which improves both detection coverage and localization precision by focusing on the first frontier that remains unmatched.

\subsubsection{Discovering New Bugs}  
  
Beyond reproducing historical bugs, we ran \appname on recent releases of all evaluated model families under both default and non-default configurations such as alternative attention backends. This process uncovered 8 previously unknown implementation issues (Table~\ref{tab:bugs-new}). Among these 8 issues, 2 have already been confirmed and patched, while the other cases have been confirmed by developers and now have follow-up fixes under review. We highlight two representative cases below.
  
\emph{GPT-2 \texttt{scale\_attn\_weights}  
  (Bugs~\#14--\#16).}  
\appname found that the \texttt{scale\_factor} parameter  
is silently ignored under the SDPA and FlashAttention  
backends in both Transformers and vLLM.  
In our framework, the candidate rule for the corresponding attention computation fails validation because the scaling factor is absent from one side. This bug is configuration-sensitive and would not surface under the default backend.
Fixes have been merged in Transformers  
(\newhref{\HF/pull/44397}{PR\#44397})  
and are under review in vLLM  
(\newhref{\VL/pull/35436}{PR\#35436}).  
  
\emph{Qwen1 \texttt{logn\_attn} (Bug~\#17).}  
Qwen1 applies a log-$n$ attention scaling during long-context inference, but this logic was entirely missing in vLLM. \appname reported non-equivalence along the attention path and localized the discrepancy to the missing scaling behavior. Unlike configuration-specific bugs that only surface under non-default settings, this bug can already be triggered under the default configuration once the input length exceeds the base context window.

  
  

\subsection{RQ2: Equivalence Proving and Scalability}\label{subsec:rq2}  

In this section, we evaluate \appname's ability to establish full equivalence for correct implementation pairs and analyze how verification time scales with model size.

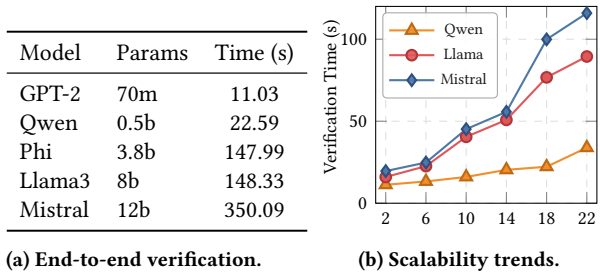
\begin{figure}
  \begin{subfigure}[b]{0.4\columnwidth}  
    \begin{tabular}{llc}  
      \toprule  
      Model   & Params & Time (s) \\  
      \midrule  
        GPT-2   & 70m   & 11.03 \\
  Qwen    & 0.5b  & 22.59 \\
  Phi     & 3.8b  & 147.99 \\
  Llama3  & 8b    & 148.33 \\
  Mistral & 12b   & 350.09 \\
      \bottomrule  
    \end{tabular}  
    \caption{End-to-end verification.}  
    \label{fig:equiv-table}  
  \end{subfigure}\hfill
  \begin{subfigure}[b]{0.58\columnwidth}  
    \centering  
    \begin{tikzpicture}
      \begin{axis}[  
        width=0.92\linewidth,  
        height=4.2cm,  
        xmin=1, xmax=23,  
        ymin=0, ymax=120,  
        xtick={2,6,10,14,18,22},  
        ylabel={Verification Time (s)},  
        ylabel style={xshift=3pt, yshift=-6pt},
        legend pos=north west,  
        legend style={  
          font=\scriptsize,  
          fill=white,  
          fill opacity=0.9,  
          draw=black!35,  
          rounded corners=1.5pt,  
          inner sep=2pt,  
          row sep=0pt,  
        },  
        grid=major,  
        major grid style={dashed, gray!25},  
        tick label style={font=\footnotesize},  
        label style={font=\footnotesize},  
        every axis plot/.append style={thick},  
      ]  
      \addplot[  
        color={rgb,255:red,242;green,142;blue,43},  
        mark=triangle*,  
        mark size=2.6pt,  
        mark options={  
          fill={rgb,255:red,242;green,142;blue,43},  
          draw={rgb,255:red,190;green,105;blue,20},  
        },  
      ] coordinates {  
        (2,11.25) (6,13.28) (10,16.02) (14,20.36) (18,22.24) (22,33.85)  
      };  
      \addlegendentry{Qwen}  
  
      \addplot[  
        color={rgb,255:red,225;green,87;blue,89},  
        mark=*,  
        mark size=2pt,  
        mark options={  
          fill={rgb,255:red,225;green,87;blue,89},  
          draw={rgb,255:red,175;green,55;blue,58},  
        },  
      ] coordinates {  
        (2,15.99) (6,22.61) (10,40.47) (14,50.81) (18,76.74) (22,89.46)  
      };  
      \addlegendentry{Llama}  

      \addplot[
        color={rgb,255:red,78;green,121;blue,167},
        mark=diamond*,
        mark size=2.2pt,
        mark options={
          fill={rgb,255:red,78;green,121;blue,167},
          draw={rgb,255:red,45;green,80;blue,130},
        },
      ] coordinates {
        (2,19.59) (6,24.82) (10,45.16) (14,55.71) (18,99.90) (22,115.90)
      };
      \addlegendentry{Mistral}
      \end{axis}  
    \end{tikzpicture}  
    \caption{Scalability trends.}  
    \label{fig:scale}  
  \end{subfigure}
  \caption{RQ2 results on correctness and scalability.  
           }  
  \Description{RQ2 results with two subfigures. The left subfigure is a table of end-to-end verification times for five model families: GPT-2, Qwen, Phi, Llama3, and Mistral. The right subfigure is a line chart showing verification time increasing with layer count for Qwen, Llama, and Mistral, with Llama and Mistral growing more steeply than Qwen.}
  \label{fig:rq2-results}  
\end{figure}

Our benchmark covers five representative model families:   
GPT2~\cite{gpt2}, Qwen~\cite{qwen}, Llama~\cite{llama3},   
Mistral~\cite{jiang2023mistral7b}, 
and Phi~\cite{abdin2024phi3technicalreporthighly},   
which are among the most widely adopted open-weight language models in the research community.  
Figure~\ref{fig:equiv-table} reports the end-to-end verification time for each model family, together with the corresponding model size in parameters.  
Across these model families, the measured verification time ranges from 11.03 seconds for GPT-2 70M to 350.09 seconds for Mistral 12B.  
For every model in the benchmark, \appname establishes that the output nodes of the two implementations belong to the same equivalence class, confirming full functional equivalence.  
Disabling any one of the key optimizations described in \S\ref{subsec:efficiency-opt} caused verification to exceed one hour on the complete benchmark. This indicates that these optimizations are essential to making the approach practical, rather than merely improving constant factors.

Figure~\ref{fig:scale} shows the scalability trends on three representative model families: Qwen, Llama, and Mistral. As shown in the figure,
verification time increases with model depth and overall architectural complexity. Qwen grows relatively steadily over the evaluated range, whereas Llama and Mistral increase much more sharply at larger depths, with Mistral exceeding 100 seconds at 22 layers. 
An explanation is the growth of the candidate search space: after dependency- and value-based filtering, \appname still needs to consider many possible relations among surviving nodes, and the number of such candidate pairs can grow quadratically in the worst case. 
As a result, deeper and more structurally diverse models require more verification work as depth increases. Despite this increase, \appname remains practical on the evaluated models.

\begin{figure*}[!t]
\begin{tcolorbox}[
  title={\appname (Ours): Automatically Synthesized Rules},
    colback=gray!5!white, colframe=gray!60!black,
  fonttitle=\bfseries\small, boxrule=0.6pt, left=1pt, right=1pt, top=0pt, bottom=0pt,
  before upper={\setlength{\abovedisplayskip}{2pt}\setlength{\belowdisplayskip}{2pt}\setlength{\abovedisplayshortskip}{2pt}\setlength{\belowdisplayshortskip}{2pt}}
]
\[
\hspace*{2em}\makebox[\dimexpr\linewidth-3em\relax][l]{$\displaystyle
\begin{aligned}
&\text{(R1)}\quad
\op{linear}(
  \op{gelu}\big(\op{linear}(x,\;\op{concat}(b_1,b_2,0),\;\op{concat}(c_1,c_2,0))\big),\;
  \op{concat}(f_1,f_2,1),\;
  e
) 
 \\[-2pt]
&\Longleftrightarrow\op{reduce\_add}(
  \op{linear}\big(\op{gelu}(\op{linear}(x,b_1,c_1)),\;f_1,\;e\big),\;
  \op{linear}\big(\op{gelu}(\op{linear}(x,b_2,c_2)),\;f_2,\;0\big)
) \\[-2pt]
&\text{(R2)}\quad  
\op{linear}\big(\op{attention}_{\mathrm{head}}(q,k,v),\;\op{concat}(w_o^{(1)},w_o^{(2)},1),\;b\big)  \\
&\Longleftrightarrow
\op{reduce\_add}(  
  \op{linear}\big(\op{attention}_{\mathrm{head}}(q_1,k_1,v_1),\;w_o^{(1)},\;b\big),\;  
  \op{linear}\big(\op{attention}_{\mathrm{head}}(q_2,k_2,v_2),\;w_o^{(2)},\;0\big)  
)  
\end{aligned}
$}
\]
\end{tcolorbox}
\begin{tcolorbox}[
  title={Entangle (Baseline): Manually Defined Rules},
  colback=gray!5!white, colframe=gray!60!black,
  fonttitle=\bfseries\small, boxrule=0.6pt, left=1pt, right=1pt, top=0pt, bottom=0pt,
  before upper={\setlength{\abovedisplayskip}{2pt}\setlength{\belowdisplayskip}{2pt}\setlength{\abovedisplayshortskip}{2pt}\setlength{\belowdisplayshortskip}{2pt}}
]
\[
\hspace*{2em}\makebox[\dimexpr\linewidth-3em\relax][l]{$\displaystyle
\begin{aligned}
&\text{(E1)}\quad\op{addmm}\big(\op{concat}(b_1,b_2,0),\,x,\,\op{concat}(W_1,W_2,1),\,\beta,\,\alpha\big)
\;\Longleftrightarrow\;
\op{concat}\big(\op{addmm}(b_1,x,W_1,\beta,\alpha),\;\op{addmm}(b_2,x,W_2,\beta,\alpha),\,1\big) \\
&\text{(E2)}\quad\op{gelu}\big(\op{concat}(t_1,t_2,d)\big)
\;\Longleftrightarrow\;
\op{concat}\big(\op{gelu}(t_1),\,\op{gelu}(t_2),\,d\big) \\
&\text{(E3)}\quad\op{matmul}\big(\op{concat}(x_1,x_2,1),\,\op{concat}(W_1,W_2,0)\big)
\;\Longleftrightarrow\;
\op{reduce\_add}\big(\op{matmul}(x_1,W_1),\,\op{matmul}(x_2,W_2)\big) \\
&\text{(E4)}\quad\op{attention}_{\mathrm{head}}\big(
\op{concat}(q_1,\ldots,q_p,1),\,
\op{concat}(k_1,\ldots,k_m,1),\,
\op{concat}(v_1,\ldots,v_m,1)\big) \\
&\qquad\Longleftrightarrow\;
\op{concat}\big(
\op{attention}_{\mathrm{head}}(q_1,k_{\pi(1)},v_{\pi(1)}),\,
\ldots,\,
\op{attention}_{\mathrm{head}}(q_p,k_{\pi(p)},v_{\pi(p)}),\,1\big)
\end{aligned}
$}
\]
\end{tcolorbox}
  \caption{Comparison of tensor-parallel equivalence rules. Top: normalized block-level rules synthesized by \appname from execution traces. Bottom: manually authored component rules used by Entangle. All rules assume compatible tensor shapes and partition layouts. R2 and E4 additionally assume that query, key, and value tensors are partitioned into disjoint head groups under a fixed query-to-KV head mapping $\pi$.}
  \Description{A comparison of tensor-parallel equivalence rules. The top panel shows two higher-level block rules synthesized by Emerge for feed-forward and head-partitioned attention computations. The bottom panel shows four lower-level manually authored Entangle rules for affine decomposition, activation distribution, reduction, and attention partitioning.}
  \label{fig:parallel-rules-compare}
\end{figure*}

\subsection{RQ3: Rule Characterization }\label{subsec:rq3}  
  
We characterize the rules synthesized by \appname from two perspectives: the distinct rule patterns that emerge across all experiments, and the granularity of these synthesized rules relative to manually authored rules.

\subsubsection{Synthesized Rule Catalogue}

Across the five full-model positive examples in our benchmark, \appname synthesizes 86 rule instances, corresponding to only
54 unique rules. Each model requires a small learned rule set, ranging from 14 to 21 rules. 
This indicates that the checking process relies not on a large model-specific rule library, but on a compact reusable core plus a few model-specific rules.
The learned rules exhibit a clear reuse pattern. 12 of the 54 distinct rules appear in at least two models, and several
recur in all models, including wrapper elimination, dtype normalization, and simple algebraic normalization. The most common
rule families are layout/wrapper normalization, normalization-path canonicalization, and partition/slicing rules. In contrast, model-
specific differences are concentrated in a small number of higher-value bridge rules, especially for attention kernels, embedding, and fused MLP.

  Of the 86 rule instances, 64 are \emph{formally verified} and 22 are \emph{empirically validated}. Most rules over explicit tensor transformations, such as
  layout, slicing, and normalization rewrites, can be discharged by SMT. By contrast, rules involving opaque or fused operators, especially attention-related
  ones, are validated through randomized testing. This division reflects the design of \appname: use formal proof whenever the operator semantics are explicit,
  and fall back to empirical validation for semantically opaque kernels. In our benchmark, we did not observe any empirically validated rule that later led to an incorrect merge or an incorrect equivalence result. Nevertheless, these rules remain testing-backed rather than formally proved.
  
\subsubsection{Comparison with Manually Authored Rules}  
To assess rule quality, we compare the rules synthesized by \appname with manually authored rules from Entangle~\cite{wang2025verifydistributeddeeplearning} on tensor-parallel equivalence patterns.
Although \appname is designed for cross-framework equivalence, this setting provides a direct way to contrast automatically inferred rules with hand-engineered ones.

Figure~\ref{fig:parallel-rules-compare} shows that \appname synthesizes block-level rules (R1--R2) that directly capture the semantic decomposition of feed-forward and head-partitioned attention blocks across tensor-parallel partitions.
For readability, R1--R2 present normalized block-level summaries of the actual learned rules, which in the implementation also contain lower-level operators such as \texttt{view}, \texttt{chunk}, fused attention kernels, and \texttt{all\_reduce}. By contrast, Entangle uses manually authored component rules (E1--E4) for affine decomposition, activation distribution, reduction, and attention partitioning. The same block-level equivalences can be recovered by composing these component rules together with standard operator normalization and implementation-specific reasoning, whereas \appname synthesizes the composed equivalence directly.

\section{Discussion}\label{sec:discussion}

In this section, we discuss the threats to validity and limitations of our approach.

\subsection{Threats to Validity}

\textbf{Internal Validity.}
Our results depend on the correctness and completeness of the captured computation graphs. Since the current implementation relies on TorchDynamo~\cite{torchfx}, bugs or omissions in graph capture may hide behavior outside the extracted graph and thereby affect both bug detection and equivalence checking. A second threat comes from rule validation. While some synthesized rules are formally proved by SMT, others are accepted through constraint-aware randomized testing. Such empirically validated rules may still be unsound, potentially leading to false equivalence.
Some results may also be sensitive to hyper-parameters such as numerical tolerances, iteration limits, and validation budgets.

\textbf{External Validity.}  
Our evaluation focuses on transformer-based language models implemented in HuggingFace Transformers and vLLM~\cite{wolf-etal-2020-transformers,vllm}. The results may not directly generalize to other serving or compilation ecosystems such as TensorRT, XLA, or ONNX Runtime~\cite{xla2017,onnx,mlir}, where operator lowering and execution behavior can differ.
Although the underlying formulation is not specific to these two frameworks,  our current implementation relies on TorchDynamo~\cite{torchfx} for computational graph capture, which may not be available or equally effective  in other deployment settings.  

\subsection{Limitations}  
  
  \textbf{Coverage and Dynamic Control Flow.}
Our approach is execution-driven: equivalence relations and rewrite rules are synthesized only for computation paths exercised by the input set. If a data-
dependent branch is never triggered, the corresponding semantics remain unexplored. This limitation is especially pronounced for Mixture-of-Experts
architectures~\cite{moe}, where expert routing depends on input values, and for long-context mechanisms such as LongRoPE, which are activated only beyond
certain sequence-length thresholds.

\textbf{Validation Strength.}
When both sides of a rule consist of symbolically tractable operators, \appname can discharge equivalence through SMT solving. However, opaque or fused kernels
often lack accessible symbolic semantics and therefore fall back to constraint-aware randomized testing. In such cases, confidence depends on the quality of
inferred preconditions and the testing budget, rather than on a formal proof.

\textbf{Scalability.}
Repeated graph execution, e-graph rebuilding, and rule enumeration introduce overhead that grows with model size. Although we employ pruning and bounded enumeration, synthesis can still become costly on very large graphs with high branching factors. We have not yet evaluated \appname at the 200B-parameter scale,
and reducing this overhead remains an important direction for future engineering work.

\section{Related Work}\label{sec:related}


\subsection{Equivalence Checking for DL Systems}

  The closest line of work studies equivalence checking for distributed training and parallel implementations of deep learning systems~\cite{trainverify, TTrace,
  Aerify, wang2025verifydistributeddeeplearning, zulkifli2025verifyingcomputationalgraphsproductiongrade}, which can be broadly grouped into three
  categories.

\textbf{Differential Testing and Diagnosis.} Testing-based approaches validate correctness by comparing executions against a trusted reference~\cite{TTrace}. Rather than comparing only final outputs, TTrace aligns intermediate tensors from a distributed execution with those from a reference implementation and uses tolerance analysis to distinguish bug-induced deviations from numerical error. However, such methods still provide empirical rather than formal evidence of equivalence. They are also sensitive to implementation differences that make numerical alignment difficult, even when the two implementations are semantically equivalent.

\textbf{Formal Verification of Parallel Training.}
TrainVerify~\cite{trainverify} targets formal verification of distributed LLM training. Given a logical model specification as the ground truth, it verifies that a distributed parallel execution plan is mathematically equivalent to that specification. To scale to frontier models, TrainVerify introduces a stage-wise verification algorithm together with shape-reduction techniques. Compared with our setting, TrainVerify focuses on parallelization equivalence within distributed training, whereas we target cross-framework model implementations whose divergences often involve different operator vocabularies, graph structures, layout transformations, and opaque kernels.

\textbf{Rewrite-based Equivalence Checking.}
Recent work~\cite{Aerify,zulkifli2025verifyingcomputationalgraphsproductiongrade,wang2025verifydistributeddeeplearning} has explored the use of equality saturation and e-graphs to check equivalence between computation graphs. Aerify~\cite{Aerify}, Entangle~\cite{wang2025verifydistributeddeeplearning}, and Scalify~\cite{zulkifli2025verifyingcomputationalgraphsproductiongrade} are representative systems in this direction. In particular, Scalify uses bijection inference to reason about layout transformations, while our work also targets equivalences arising from different operator implementations and execution structures.

Compared with these approaches, our work focuses on discovering rules dynamically during verification. This makes the approach better suited to modern large-model implementations, where the relevant transformations are often framework-specific and difficult to enumerate in advance.

\subsection{Equality Saturation and Rule Synthesis}

Our approach builds directly on e-graphs and equality saturation~\cite{equalitysaturation}, which provide a compact representation of many equivalent
expressions and support efficient congruence maintenance during rewriting. Egg~\cite{egg} introduces an efficient equality-saturation engine together with the
rebuild procedure that restores congruence after merges; our rebuilding procedure is inspired by this design. Our attachment of concrete execution values to e-classes can be viewed as an e-graph analysis in the sense of Egg~\cite{egg}. Egglog~\cite{egglog} further combines equality saturation with Datalog-style reasoning to improve rule application efficiency.
Beyond applying a fixed rewrite set, several prior works study the automatic inference of rewrite rules. Ruler~\cite{ruler} and Enumo~\cite{enumo}, for example,
synthesize rules from a grammar and an interpreter by enumerating candidate expressions and validating semantic equivalence. 
TASO~\cite{TASO} and Mirage~\cite{mirage} are also related in that they enumerate graph substitutions and verify their correctness under known operator semantics, but their goal is graph optimization rather than implementation equivalence.

Our work differs in two important ways. First, we synthesize rules in the context of large deep learning computation graphs rather than in a standalone algebraic domain. Second, our synthesis procedure is guided by the shared-frontier structure induced by the current e-graph, which constrains the search space and enables practical rule
discovery on real models.

\section{Conclusion}\label{sec:conclusion}

In this paper, we presented \appname, a verification engine for \emph{Implementation Equivalence}. Instead of relying on manually written rewrite rules, \appname infers candidate relations from execution values, synthesizes rules on demand, validates them, and propagates through a joint e-graph.
  Our current implementation and evaluation focus on inference computation graphs, and the results show that this execution-driven design is effective in practice: \appname detects and localizes real implementation bugs, establishes equivalence for correct model pairs at practical cost, and synthesizes rules that compare with manually authored ones. 


\bibliography{references}

\end{document}